\documentclass[12pt]{article}

\usepackage{secdot}

\def\qed{\relax\ifmmode\hskip2em \fbox{ }\else\unskip\nobreak\hskip1em 
$\fbox{}$\fi}

\newsavebox{\theorembox}
\newsavebox{\lemmabox}
\newsavebox{\corollarybox}
\newsavebox{\propositionbox}
\newsavebox{\examplebox}
\newsavebox{\propertybox}
\savebox{\theorembox}{\bf Theorem}
\savebox{\lemmabox}{\bf Lemma}
\savebox{\corollarybox}{\bf Corollary}
\savebox{\propositionbox}{\bf Proposition}
\savebox{\examplebox}{\bf Example}
\savebox{\propertybox}{\bf Property}
\newtheorem{theorem}{\usebox{\theorembox}}

\newtheorem{definition}{{\sc Definition}\rm }[section]
\newtheorem{definitions}[definition]{{\sc Definitions\rm }}

\newenvironment{proof}{{\noindent\bf Proof~}}{\(\qed\)\vspace*{\proofskip} }
\newlength{\proofskip}
\begin{document}

\begin{center}

{\LARGE 
An Optimal Offline Algorithm for List Update
}


\footnotesize

\mbox{\large Srikrishnan Divakaran}\\
DA-IICT, 
Gandhinagar, Gujarat, India 382007\\
\mbox{
Srikrishnan\_divakaran@daiict.ac.in }\\[6pt]
\normalsize
\end{center}

\baselineskip 20pt plus .3pt minus .1pt


\noindent
\begin{abstract}
For the static         list update problem, given an ordered list $\rho_0$ (an ordering of    the list $L$ = \{ $a_a, a_2, ..., a_l$ \}), and a sequence $\sigma = (\sigma_1,        \sigma_2, ..., \sigma_m)$ of requests for items in $L$, we   characterize the list reorganizations in an optimal offline solution in terms of an initial permutation of the list     followed by a sequence of $m$    
{\em element transfers}, where an  element   transfer is a type of list reorganization where only the requested item can  be moved. Then we make use of this characterization to design an $O(l^{2}
(l-1)!m)$ time optimal offline algorithm.
\end{abstract}
\noindent\hrulefill
\section{Introduction}
A dictionary is   an abstract data type that stores a collection of distinct items and supports the operations    {\em access,   insert, and delete} based on their key values. In {\em the list update problem}[8], the dictionary is implemented as a simple linear list $L$ where the items are stored as a linked collection of items. The cost of servicing a   request for an item $a \in L$ is 1 plus the number of   items   preceding $a$ in the list. That is, accessing or deleting the $i$th item of $L$ costs $i$. Inserting   a   new   item costs $l+1$, where $l$ is the number of items in $L$ prior to insertion. For any given sequence of requests for access, insert or delete of items of list $L$, an algorithm   may reorganize $L$ from  time to time in an attempt to reduce the access cost of future requests. The list reorganization is done using sequence of transpositions of   consecutive  items. If  the list reorganization involves moving   the most recently accessed item forward then we refer to the transpositions involved in this reorganization to be {\em free transpositions},    otherwise  we refer to them as {\em paid transpositions}. The cost for reorganizing a list is usually measured in terms of the   number of        paid transpositions  it uses. More formally, we  can define {\em the list update problem} as follows:
\begin{quote}
Given an     ordered list  $\rho_0$ (an ordering among the items in $L$) and  a request     sequence $\sigma$, we need to determine  how to reorganize  $L$ while serving  $\sigma$ so as to minimize the total servicing (access and list reorganization) cost. 
\end{quote}
This    problem        is usually referred to as {\em the dynamic list update problem}[8]. There is a simpler version of this problem, referred to as {\em the static list update problem}[8], where the list consists of a  fixed set of $l$ items and insertions or deletions are not allowed. For many standard cost models, the static    list   update        and the dynamic list update problems are known to be equivalent.  In the list update problem, if we have knowledge of the complete request sequence prior to   start   of   servicing  then we  refer to this problem as {\em the offline list update problem}[8]. However, at any time  if the current request has to be served with no knowledge of future requests then we refer to this problem as the {\em the online list update problem}[8]. \newline \newline
In this paper, our focus is on the design of efficient offline algorithms for the static list update problem. For the static list update problem, servicing requests offline essentially     involves (i) reorganizing the list either before or after accessing the requested item, and (ii) accessing    the requested item. Traditional offline          algorithms[8] for list update reorganize their lists using subset transfers. In a subset  transfer, the   list is     reorganized prior to access by  moving  a subset $S$ of items that are in       front of the requested item   while  maintaining the  relative 
ordering of items of $S$. We     however  present offline algorithms  that except for its first list reorganization uses only {\em element transfers}. In     an    element transfer,  after    an access the list can  be reorganized by moving only the requested   item to any position (forward as well as backward) in   the list. In both       these   types of list         reorganizations, we  define the reorganization cost to be the number of     transpositions used instead of only considering      the number of paid transpositions. This way of accounting the reorganization cost helps      to keep our algorithm and its analysis simple without increasing the overall servicing cost.       \footnote{The optimal       cost of servicing a request sequence is the same irrespective of whether the reorganization cost is measured in terms of number of paid transpositions used     or    in terms of number of transpositions (free or paid) used.} \newline \newline
{\bf Related Results}: Algorithms    for      both offline and online list update problems have been investigated by many researchers. For a comprehensive  study of these algorithms we refer the reader to [l-4, 6-9, 11-14]. In this paper, our focus   is     on offline algorithms for the static list update problem. For the offline list update problem, Reingold and Westbrook[12] characterized  optimal solutions in terms of subset transfers, and used this characterization to present an $O(2^{l}(l-1)! 
m)$ time and $O(l!)$ space optimal  offline algorithm. Then, Pietzrak[10] presented an $O(l^{3}l!
m)$ time forward   dynamic  program by  making use of the observation that       any     subset transfer involves at most $l(l-1)/2$ consecutive transpositions. Recently, Ambuhl[5] showed this problem to be NP-hard. \newline \newline
{\bf Our Results}: We characterize the  list reorganizations in an optimal offline solution in terms of an initial permutation of the list   followed by a sequence of $m$ {\em element transfers}, where an element transfer is a type of list     reorganization where only the requested item can be moved.
Then, we    make use of this simpler characterization to design  an   $O(l^2(l-1)!m)$ time  optimal offline algorithm for list update.  \newline \newline
{\bf Paper Outline}: The rest of this paper is organized as follows. In Section $2$, we present  the characterization of optimal offline solutions  for list update in terms of (a) subset  transfers and 
(b) element transfers. In  Section $3$,  we make use    of the characterization in terms  of element transfers to present an $O(l^2(l-1)!m)$ time optimal offline Algorithm for list update.
\section{Characterization of Optimal Offline Solutions for List Update }
In Section $2.1$, we first introduce  terms    and definitions necessary for     characterizing list reorganizations in optimal offline solutions for  list update. Then, in Section $2.2$, we    present Reingold and Westbrook's[12] characterization of optimal offline  solutions for list update   in terms of a sequence of subset transfers. Finally, in Section $2.3$,  we present our    characterization of optimal offline solutions for list update in  terms of an   initial permutation of the list followed by a sequence of element transfers.
\subsection{Basic Terms and Definitions}
\begin{definitions}
Let $L$ = \{ $a_1,  a_2, ..., a_l$ \} be a list of distinct items, $\cal{P}$ be   the set of all orderings of the items in $L$. We define
\begin{itemize}
\item[-] $\rho \in \cal{P}$  to be   an ordering of the items in $L$;
\item[-] $pos^{a}(\rho)$   be the position of item $a$ in  $\rho$;
\item[-] $\rho[i..j]$ to    be the ordered sub-list consisting  of the items in $\rho$ starting 
at position $i$ and ending at position $j$.
\end{itemize}
\end{definitions}
\begin{definitions}
Let $\rho \in \cal{P}$ be an ordered list and $a \in L$ be an  item at position $k$   in $\rho$. We define
\begin{itemize}
\item [-] $ST^{a}_{S}(\rho)$, a subset transfer in $\rho$  with respect to the item $a$ and  set $S \subseteq \rho[1..k-1]$, as  a minimal set of transpositions of consecutive items          used  to reorganize $\rho$ by moving all the items of $S$ to the right of $a$;
\item[-] $config^{a}_{S}(\rho)$ to be the list configuration that results after the subset transfer 
$ST_{S}^{a}(\rho)$;
\item[-] $cost^{a}_{S }(\rho) = |ST^{a}_{S}(\rho)|$ to be the cost associated     with the   subset  transfer $ST_{S}^{a}(\rho)$ measured in terms of the number of transpositions used in $ST_{S}^{a}(\rho)$;
\item[-] $ST^{a}(\rho)$ = \{ $ST_{S}^{a}(\rho) : S \subseteq \rho[1..k-1]$  \} to   be the   set of all subset transfers with respect to item $a$.
\end{itemize}
\end{definitions}
{\bf Note}: Since item $a$ is at position $k$ in $\rho$,    there are $2^{k-1}$ distinct subsets of  
$\rho[1..k-1]$. Hence                 $ST^{a}(\rho)$ consists of       $2^{k-1}$  different  subset  transfers.   Once  the set $S \subseteq \rho[1..k-1]$ is  specified, the subset transfer  in $\rho$ with respect to the item $a$ and set $S$  corresponds to an unique set of consecutive transpositions and  the relative ordering of items of  $S$ in $\rho$ remains unaffected during the subset transfer. 
\begin{definitions}
Let $\rho$ be an ordered list and $a\in L$   be an item at position        $k$ in $\rho$. We define
\begin{itemize}
\item[-] $ET_{j}^{a}(\rho)$, an element transfer in  $\rho$ with respect to item $a$ and an integer position $j \in [1..l]$, to be the minimal set of consecutive transpositions for reorganizing $\rho$ such  that $a$ is moved to position $j$ in the list;
\item[-]  $config^{a}_{j}(\rho)$ to be  the    ordered list that results after the element transfer 
$ET_{j}^{a}(\rho)$;
\item[-] $cost^{a}_{j}(\rho) =|ET_{j}^{a}(\rho)|$ to be the cost        associated with the element transfer $ET_{j}^{a}(\rho)$ measured   in terms of the number of transpositions used in $ET_{j}^{a}
(\rho)$;
\item[-] $ET^{a}(\rho)$ = \{ $ET_{j}^{a}(\rho) : j \in [1..l]$ \} to be the    set of all  element transfers with respect to item $a$.
\end{itemize}
\end{definitions}
{\bf Note}: We are      allowed to move the requested item to any position in the list. Therefore,  there are  $l$ different  element transfers possible   with respect to the requested item and also the  relative ordering of    items other than the requested item are unaffected during the element transfer.
\subsection{Characterization of Optimal Offline Solutions in terms of subset transfers}
Given an ordered list $\rho_0$ and an    arbitrary request sequence $\sigma = (\sigma_1, \sigma_2, ..., \sigma_m)$, Reingold   and Westbrook[12] established that  an optimal offline solution for list 
update  can be obtained by reorganizing the list using a sequence of $m$ subset transfers.    More  formally, we  only need to consider offline algorithms that for $i \in [1..m]$ services $\sigma_i$
by reorganizing  its list using a subset transfer with respect to item  $\sigma_i$ and        then
access      $\sigma_i$.   We   now present some definitions     that  help us present Reingold and Westbrook's characterization of optimal offline solutions for list update.
\begin{definitions}
Let $\rho_0 \in \cal{P}$ be an ordered list on $L$, $\sigma = (\sigma_1, \sigma_2, ..., \sigma_m)$ be an arbitrary sequence of requests for items in $L$, and $A$ be an  offline   algorithm for list update that reorganizes its list using only subset transfers. We now define  
\begin{itemize}
\item[-] $A(\sigma) = (A_1(\sigma), A_2(\sigma), ..., A_{m}(\sigma))$ to be the sequence of subset 
transfers performed while servicing $\sigma$;
\item[-] For $i \in [1..m]$, 
\begin{itemize}
\item[-] $A_{i}^{a}(\sigma)$, $a \in L$, to be the transpositions involving   $a$ in $A_{i}(\sigma)$;
\item[-] $\rho^{A}_{i}$, to be $A$'s list configuration after the subset transfer $A_{i}(\sigma)$;
\item[-] $Trans^{A}(\sigma_i)=|A_{i}(\sigma)|$      to be $A$'s rearrangement  cost 
while  servicing $\sigma_i$;
\item[-] $Access^{A}(\sigma_i) = pos^{\sigma_i}(\rho^{A}_i)$ to be $A$'s  cost for accessing $\sigma_i$;
\item[-] $Cost^{A}(\sigma_i) = Trans^{A}(\sigma_i) + Access^{A}(\sigma_i)$ to be $A$'s cost for servicing $\sigma_i$.
\end{itemize}
\end{itemize}
\end{definitions}
\begin{theorem}
For the Static List Update Problem, given           an ordered list   $\rho_0 \in \cal{P}$ on $L$, and a sequence $\sigma = (\sigma_1, \sigma_2, ..., \sigma_m)$ of requests for items in $L$,  there exists an optimal offline solution where $\sigma$ is serviced by reorganizing its list     using a sequence of    $m$ subset transfers.
\end{theorem}  
We refer the reader to the papers of Reingold and Westbrook [12] for the proof of Theorem $1$.
\subsection{Characterization of Optimal Offline Solutions in terms of element transfers}
Given an ordered list $\rho_0$ and an arbitrary request  sequence $\sigma=(\sigma_1, \sigma_2, ..., \sigma_m)$, we show that there exists an optimal offline solution for list update    where the list reorganization is done by permuting $\rho_0$       followed by   a sequence          of $m$ element transfers.  More formally, we only      need to  consider offline algorithms that service  $\sigma$ by first permuting $\rho_0$ prior to servicing $\sigma$, and  then for $i \in [1..m]$,   services $\sigma_i$  by    reorganizing          its list      using an  element transfer with   respect to $\sigma_i$, and then access   $\sigma_i$. We  now present some definitions that will help us present our    characterization of optimal offline solutions for list update.
\begin{definitions}
Let $\rho_0$ be an ordered list on $L$, $\sigma = (\sigma_1, \sigma_2, ..., \sigma_m)$ be a sequence of requests for items in $L$, and  $A$ be an offline   algorithm for list update that prior to start of servicing permutes its list $\rho_0$ and then services  $\sigma$ by using a       sequence of $m$ element transfers. We   now define 
\begin{itemize}
\item[-] $A(\sigma)= (A_0(\sigma), A_1(\sigma), A_2(\sigma), ..., A_{m}(\sigma))$ to be the sequence of list rearrangements performed while         servicing $\sigma$, where $A_0(\sigma)$ is the set of consecutive transpositions used in permuting $\rho_0$, and  $A_{i}(\sigma)$, for $i \in [1..m]$,  be the element transfer performed by $A$ while servicing $\sigma_i$;
\item[-] For $i \in [1..m]$,  
\begin{itemize}
\item[-] $A_{i}^{a}(\sigma)$,  for $a \in L$,   to  be the   transpositions involving $a$ in $A_{i}(\sigma)$;
\item[-] $\rho^{A}_{i}$, to be $A$'s list configuration after $A_{i}(\sigma)$;
\item[-] $Trans^{A}(\sigma_i)=|A_{i}(\sigma)|$ to be $A$'s rearrangement cost while     servicing $\sigma_i$;
\item[-] $Access^{A}(\sigma_i) = pos^{\sigma_i}(\rho^{A}_i)$ to be $A$'s cost for accessing 
$\sigma_i$;
\item[-] $Cost^{A}(\sigma_i) = Trans^{A}(\sigma_i) + Access^{A}(\sigma_i)$    to be $A$'s cost  for servicing $\sigma_i$.
\end{itemize}
\end{itemize}
\end{definitions}
\begin{theorem}
For the Static List Update Problem, given an ordered list   $\rho_0 \in    \cal{P}$ on  $L$, and a sequence $\sigma= (\sigma_1, \sigma_2, ..., \sigma_m)$ of requests for items in $L$,  there exists 
an optimal       offline solution where $\rho_0$ is permuted first    and then         $\sigma$ is  serviced  using a sequence of $m$ element transfers.
\end{theorem}
Now, we introduce certain terms that we find convenient in proving Theorem $2$.
\begin{definitions}
Let $\sigma = (\sigma_1,\sigma_2,...,\sigma_m)$ be a sequence of requests for       items in $L$.
We define
\begin{itemize}
\item[-] $first^{a}(\sigma)$ to be the position in $\sigma$ of  the first occurrence of  request for item $a$; 
\item[-] $next^{a}(\sigma_i)$ to be the position in $\sigma$ of the first request to item $a$  after  $\sigma_i$.
\end{itemize}
\end{definitions}
\begin{proof}
{\bf of Theorem $2$}: 
Let $OPT$ be an          optimal offline algorithm for the list update problem. From  Reingold and Westbrook's characterization, we know that there exists an optimal offline solution $OPT(\sigma) = 
(OPT_{1}(\sigma), OPT_{2}(\sigma), ..., OPT_{m}(\sigma))$, where for $i \in [1..m]$,     $OPT_{i}(\sigma)$, are subset transfers with respect to $\sigma_i$. Now, from $OPT(\sigma)$ we    construct an offline        solution $B(\sigma) = (B_0(\sigma), B_1(\sigma), ..., B_m(\sigma))$,       where 
$B_0(\sigma) = \bigcup_{i=1}^{l} \bigcup_{j=1}^{first(a_i)} OPT^{a_i}_{j}$ is      a permutation of 
$\rho_0$, and for $i \in [1..m]$, $B_{i}(\sigma)=\bigcup_{j=i+1}^{next(\sigma_i)} OPT^{\sigma_i}_{j
}$ is an element transfer with respect to $\sigma_i$. We will now show that  $B(\sigma)$ is also an optimal solution for $\sigma$. \newline \newline
From the construction of $B(\sigma)$, we can observe that    the transpositions used in $B(\sigma)$ are the same as in $OPT(\sigma)$, so the total reorganization cost in $B(\sigma)$ is the same as in 
$OPT(\sigma)$. Therefore, to prove that $B(\sigma)$ is also optimal it is sufficient to  show that
for $i \in [1..m]$, $Access^{B}(\sigma_i) = Access^{OPT}(\sigma_i)$. \newline \newline
Let $i \in [1..m]$ be some arbitrary integer. From the construction of $B(\sigma)$, we can observe that  just prior to accessing $\sigma_i$, the $i$ reorganizations $B_0(\sigma), B_1(\sigma), ..., 
B_{i-1}(\sigma)$ have been performed on its list. This includes (i) all  the transpositions in the 
first $i$ subset transfers    $OPT_1(\sigma),..., OPT_i(\sigma)$ performed in $OPT(\sigma)$    and 
(ii) for each element $a \ne  \sigma_i$ in $L$, the transpositions involving   $a$ in  $OPT_{i+1}(\sigma), ..., OPT_{next^{a}(\sigma_i)}(\sigma)$. The transpositions in (i) are  common to both $B(\sigma)$ and $OPT(\sigma)$.  Therefore,  if we show that the transpositions in (ii) does       not affect the position of $\sigma_i$ in $B(\sigma)$ then we are done.  \newline \newline
Let   $a \ne \sigma_i$ be some arbitrary item in $L$. We   will show that at the time of accessing
$\sigma_i$ the transpositions involving $a$ that are done in $B(\sigma)$ but not yet done in $OPT(\sigma)$ do  not affect the position of $\sigma_i$ in  $B(\sigma)$.  Notice  that     there are no requests for $a$ between $\sigma_i$ and $next^{a}(\sigma_i)$, so all transpositions involving  $a$ in  $OPT_{i+1}(\sigma), ..., OPT_{next^{a}}(\sigma)$ will only move $a$ away from the front of the list. Now,   based   on  the relative ordering of $a$ and $\sigma_i$ in $OPT$ and $B$          the following situations are possible:
\begin{itemize}
\item[] {\bf Case $1$:   $a$ is before $\sigma_i$ in both $OPT$ and $B$}: In   this      situation  the transpositions in (ii) involving $a$ does not affect the position of $\sigma_i$ in $B$. So, we  are done.
\item[] {\bf Case $2$: $a$ is after $\sigma_i$ in both $OPT$ and $B$}: In this situation also  the transpositions in (ii) involving $a$ does not affect the position of $\sigma_i$ in $B$. So, we are 
done.
\item[] {\bf Case $3$: $a$ is before $\sigma_i$ in $OPT$ and after $\sigma_i$ in $B$}: In this case
we can make $OPT$ also perform this transposition before accessing $\sigma_i$ and lower  the  total servicing cost. This would contradict the optimality of $OPT$ and hence this      situation  is not possible. 
\item[] {\bf Case $4$: $a$ is after $\sigma_i$ in $OPT$ and before      $\sigma_i$ in $B$}:  Notice that all transpositions involving   $a$ in (ii) will only move it away from the front  of the list. So this situation is not possible.
\end{itemize}
\end{proof}
\section{Our Algorithm $A$}
Given an ordered list $\rho_0$ and an  arbitrary request sequence $\sigma=(\sigma_1, \sigma_2, ..., \sigma_m)$, we make use  of our characterization        of an  optimal offline solution in terms of element transfers     to design an $O(ml^2(l-1)!)$ time   optimal offline algorithm. Our  algorithm  determines    an optimal sequence of  list reorganizations    by first constructing a $m+2$ layered Action Network  $AN(\sigma)$ with         a source node $s$ and  a destination node   $t$, and then  determines   a shortest path between the nodes $s$ and $t$.  We  will     first describe the Action Network $AN(\sigma)$ and then present our Algorithm $A$. \newline \newline
{\bf Action Network}: Given   a sequence $\sigma = (\sigma_1, \sigma_2, ..., \sigma_m)$ of requests for items in $L$ =   \{$a_1$, $a_2$, ..., $a_l$ \},   the 
{\em Action Network} $AN(\sigma)=(N^{\sigma}, A^{\sigma})$ is a layered network consisting of $m+2$ layers. Layer $0$ consists of a single node $s$ that we refer to  as the source        node of $AN(\sigma)$, layer $m+1$ consists of a single node $t$ that we refer to as the     destination node of  $AN(\sigma)$. For $i \in [1..m]$, layer $i$ consists of $l!$ nodes  $n^{\rho}_{i}$, for $\rho   \in \cal{P}$.  For $i \in [1..m]$ and $\rho \in \cal{P}$, node  $n^{\rho}_{i}$ is associated with   the ordered list $\rho$. 
For $\rho \in \cal{P}$, there is an arc from node $s$ to a node $n^{\rho}_{1}$ in layer $1$.  For $i \in [1..m-1]$ and $\rho, \rho' \in \cal{P}$, there is an arc from   node $n^{\rho}_{i}$ in layer $i$ to node $n^{\rho'}_{i+1}$ in   layer $i+1$ if $\rho' \in       config_{j}^{\sigma_i}(\rho)$, for some $j \in [1..l]$.   That is,         $e = (n_{i}^{\rho}, n_{i+1}^{\rho'})$ is     an edge in $AN(\sigma)$ if    $\rho'$ can be obtained from $\rho$ by   performing an element transfer with respect to $\sigma_i$ and some position $j \in [1..l]$. Finally, every  node in layer 
$m$ is connected to node $t$ in layer $m+1$. 
For each arc $e = (s, \rho)$ from node $s$ to a node in layer $1$ of  $A^{\sigma}$, we associate an action $action(e)$    and     cost   $cost(e)$, where $action(e)$ is the minimum set of consecutive transpositions required to transform   $\rho_0$ to $\rho$, and  $cost(e)$     is      the number of transpositions in $action(e)$. Similarly,  for each arc $e = (n_{i}^{\rho}, n_{i+1}^{\rho'})$  from a node in   layer $i \in [1..m-1]$ to node in layer $i+1$    of $\sigma$, we define  $action(e)$ to be  the  set of transpositions in the element transfer associated with  $e$ and $cost(e)$ to be the number of   transpositions  in $action(e)$. Finally, for all arc from nodes in layer $m$ to $t$, we define $action(e) = \phi$ and $cost(e) = 0$. \newline \newline
{\bf Algorithm $A$} \newline 
{\bf Basic Idea}: Given an ordered list $\rho_0$  and a sequence $\sigma$ of $m$ requests for items  in $L$, Construct a         $m+2$ layered network $AN(\sigma)$ that represents the sequence of list reorganizations of an offline solution for list update  as a path between the  nodes $s$ and $t$ in 
$AN(\sigma)$ such      that a path from $s$ to $t$ of length  $h$ exists if and only if there is an offline algorithm that can service $\sigma$ at cost of $h$. Then, we determine the optimal solution for servicing $\sigma$  by determining the actions corresponding to a shortest length path from $s$ to $t$ in $AN(\sigma)$.
\begin{tabbing}
Inp\=uts \\
   \> $\rho_0$ : initial configuration of list $L$; \\
	 \> $\sigma$ : sequence $(\sigma_1, \sigma_2, ..., \sigma_m)$ of requests for items in $L$. \\
Out\=put \\
   \> sequence of list reorganizations performed by the algorithm while servicing $\sigma$; \\
Beg\=in\\
   \> $(1)$ \= Construct a $m+2$ layered network $AN(\sigma)= (N^{\sigma}, A^{\sigma})$ such that 
	layer $0$ consists  \\
	 \>       \> of node $s$, layer $m+1$ consists of node $t$,  and   for $i \in [1..m]$,  layer $i$ 
	consists of $l!$ \\ 
	\>        \> nodes $n^{\rho}_{i}$, where $\rho   \in \cal{P}$. Node $s$ is associated with the list $\rho_0$. For $i \in [1..m]$ and  \\
	\>        \>  $\rho \in \cal{P}$, node $n^{\rho}_{i}$ is associated with the ordered list 
	$\rho$. \\
   \> $(2)$ \= For \= $\rho \in \cal{P}$ \\
	 \>     \>    \> Add an arc $e = (s, \rho)$; \\
	 \>     \>    \> Set $action(e) = inversions(\rho_0, \rho)$ and
	                     $cost(e) 	= |inversions(\rho_0, \rho)|$; \\
	 \> $(3)$ \= For \= $i \in [1..m]$ and $\rho \in \cal{P}$ \\
	 \>       \>     \> For \= $\rho' \in ET^{\sigma_i}(\rho)$ \\
	 \>       \>     \>     \> Add an arc $e = (n^{\rho}, n^{\rho'})$; \\
	 \>       \>     \>     \> Set $action(e) =  inversions(\rho, \rho')$ and 
																	$cost(e) = |inversions(\rho, \rho')|$; \\
   \> $(4)$ \= For \= $\rho \in \cal{P}$ \\
	 \>       \>     \> Add an edge $e = (n^{\rho}_{m+1}, t)$; \\
	 \>       \>     \> Set $action(e) = \phi$ and $cost(e) = 0$ \\
	 \> $(5)$ \= Find the the shortest path $SP$ from $s$ to $t$ and print $action(e)$ for each
	$e \in SP$. \\
End.
\end{tabbing}
\begin{theorem}
Given an ordered list   $\rho_0 \in \cal{P}$ on $L$, and a sequence $\sigma = (\sigma_1, \sigma_2, ..., \sigma_m)$ of requests for items in $L$, Algorithm $A$ determines an optimal offline solution for the Static List Update in $O(ml!)$ time.
\end{theorem}
\begin{proof}
From the construction of the Action Network $AN(\sigma)$, we can observe that there is a one to one 
correspondence between a path from the  node $s$ to  node $t$ and a    sequence of list organizations of an offline algorithm that permutes its list and then       services  $\sigma$ by performing a sequence of $m$ element transfers. From   Theorem $2$, we can observe that the shortest path from $s$ to $t$ in $AN(\sigma)$ will be an optimal offline solution for $\sigma$.
Notice that  $AN(\sigma)$ is a layered network and    from each node in $AN(\sigma)$ other than $s$ there are exactly $l$ edges leaving that node.    So, if   we know the shortest path from $s$ to a 
node in layer $i$ then we can determine the shortest path to all the nodes in layer $i+1$ in $l*l!$ 
computations. Since there are $m$ layers, we can compute the shortest path from $s$ to $t$ in $O(ml
*l!)$ time.
\end{proof}
\section{Conclusions and Future Work}
We have a simple characterization of optimal      offline solutions for list update in terms of an initial list permutation followed by a sequence of element transfers. This  characterization helps in reducing the run-time complexity from previously known $O(ml^3l!)$ time to $O(mll!)$ time.   We feel that this simple characterization         can lead to computationally efficient approximation algorithms/schemes with stronger approximation guarantees. We         have developed heuristics by simplifying our Algorithm $A$ and experimentally   they yield solutions very close to the optimal. However, we   are still in the process of theoretically establishing its performance guarantee.

\end{document}